\documentclass[twocolumn]{jpsj3}
\usepackage{graphicx}
\usepackage{amsmath,amssymb}

\def\runauthor{Youichi {\sc Yanase}}

\title{ 
Ginzburg-Landau Analysis for the Antiferromagnetic Order 
in the Fulde-Ferrell-Larkin-Ovchinnikov Superconductor 
} 

\author{Youichi {\sc Yanase}$^{1,2}$\footnote{E-mail:
yanase@phys.sc.niigata-u.ac.jp} and Manfred Sigrist$^2$}

\inst{
$^{1}$Department of Physics, Niigata University, Ikarashi, 
Niigata 950-2181, Japan
\\
$^2$Theoretische Physik, ETH-Honggerberg, 8093 Zurich, Switzerland
}

\recdate{Today 2011}

\abst
{ 
 Antiferromangetic (AFM) order in the Fulde-Ferrell-Larkin-Ovchinnikov 
(FFLO) superconducting state is analyzed on the basis of the 
Ginzburg-Landau theory. 
 To examine the possible AFM-FFLO state in a heavy fermion superconductor 
CeCoIn$_5$, we focus on the incommensurate AFM order characterized by the 
wave vector $\vec{Q} = \vec{Q}_{0} \pm \vec{q}_{\rm inc}$ with 
$\vec{Q}_{0} =(\pi,\pi,\pi) $ and 
$\vec{q}_{\rm inc} \parallel [110]$ or $[1\bar{1}0]$
in the tetoragonal crystal structure. 
 We formulate the two component Ginzburg-Landau model 
to discuss two denerate incommensurate AFM states with 
$\vec{q}_{\rm inc} \parallel [110]$ and $[1\bar{1}0]$.
 Owing to the broken translation symmetry in the FFLO state, 
multiple phase diagram of single-q phase and double-q phase 
is obtained under the magnetic field along 
$[100]$ or $[010]$. 
 Magnetic properties in each phase are investigated and compared 
with the neutron scattering and nuclear magnetic resonance (NMR) 
measurements. 
 An ultrasonic measurement is proposed for a future experimental 
study on the possible AFM-FFLO state in CeCoIn$_5$.  
 The field orientation dependence of the AFM order in CeCoIn$_5$ is 
discussed. 
}

\kword
{
FFLO superconductivity, unconventional magnetism, CeCoIn$_5$ 
}

\begin{document}
\sloppy
\maketitle

\newcommand{\eli}{$\acute{{\rm E}}$liashberg }
\renewcommand{\k}{\vec{k}}
\newcommand{\kk}{\vec{k'}}
\newcommand{\q}{\vec{q}}
\newcommand{\qa}{\vec{q}_{1}}
\newcommand{\qb}{\vec{q}_{2}}
\newcommand{\Q}{\vec{Q}}
\newcommand{\Qaf}{\vec{Q}_{0}}
\newcommand{\qi}{q_{\rm inc}}
\newcommand{\qiv}{\vec{q}_{\rm inc}}
\newcommand{\qia}{\vec{q}_{\rm inc}^{\rm \,\,(1)}}
\newcommand{\qib}{\vec{q}_{\rm inc}^{\rm \,\,(2)}}
\renewcommand{\r}{\vec{r}}
\newcommand{\e}{\varepsilon}
\newcommand{\ee}{\varepsilon^{'}}
\newcommand{\ep}{\epsilon}
\newcommand{\s}{{\mit{\it \Sigma}}}
\newcommand{\Tc}{$T_{\rm c}$ }
\newcommand{\Tcf}{$T_{\rm c}$}
\newcommand{\TN}{$T_{\rm N}$ }
\newcommand{\TNz}{$T_{\rm N}^0$ }
\newcommand{\TNf}{$T_{\rm N}$}
\newcommand{\Hc}{$H_{\rm c2}^{\rm P}$ }
\newcommand{\Hcf}{$H_{\rm c2}^{\rm P}$}
\newcommand{\etal}{{\it et al.} }
\newcommand{\PRL}{Phys. Rev. Lett. } 
\newcommand{\PRB}{{\it Phys. Rev.} B } 
\newcommand{\JPSJ}{J. Phys. Soc. Jpn. } 
\newcommand{\Science}{{\it Science} } 
\newcommand{\Nature}{{\it Nature} } 
\newcommand{\qf}{\vec{q}_{\rm FFLO}}
\newcommand{\qfs}{q_{\rm FFLO}}
\renewcommand{\i}{\hspace*{0.3mm}\vec{i}\hspace*{0.6mm}}
\renewcommand{\j}{\hspace*{0.3mm}\vec{j}\hspace*{0.6mm}}
\newcommand{\Co}{CeCoIn$_5$ }
\newcommand{\Cof}{CeCoIn$_5$}
\newcommand{\va}{\vec{a}}
\newcommand{\vb}{\vec{b}}
\newcommand{\vdelta}{\vec{\delta}\hspace*{0.5mm}}
\newcommand{\vH}{\vec{H}}
\newcommand{\neel}{N$\acute{\rm e}$el }

\section{Introduction}

More than 40 years ago, Fulde and Ferrell~\cite{FF}, and 
Larkin and Ovchinnikov~\cite{LO} proposed the appearance of a
spatially modulated phase in a spin polarized superconductor. 
The original BCS theory is based on a condensate of Cooper pairs with 
vanishing total momentum. On the other hand, this 
Fulde-Ferrell-Larkin-Ovchinnikov (FFLO) phase is composed of
Cooper pairs with finite momentum. 
A spontaneous breaking of spatial symmetry is implied by 
the internal degrees of freedom of the FFLO phase 
arising from inversion or reflection symmetry. 

The search for the FFLO phase has been pursued since its proposal
with mixed success. 
Indeed it has been discussed for various systems such as 
superconductors~\cite{radovan2003,PhysRevLett.91.187004,
uji:157001,singleton2000,lortz:187002,
shinagawa:147002,yonezawa:117002}, 
cold atom gases~\cite{Partridge01272006,Zwierlein01272006,Liao2010}, 
and quark matter~\cite{casalbuoni2004}. 
Nevertheless, the discovery of a 
new superconducting phase in \Co at high magnetic fields and 
low temperatures~\cite{radovan2003,PhysRevLett.91.187004} came as
a surprise and triggered much interest, as it satisfied all 
the immediate criteria for an FFLO state, sitting in the right 
parameter space in the $H$-$T$-phase diagram~\cite{matsuda2007}.  
  
The high-field superconducting (HFSC) phase of \Co has been
examined based on the FFLO-concepts by many 
groups~\cite{matsuda2007,watanabe2004,capan2004,martin2005,
mitrovic2006,miclea2006,correa2007,adachi2003, ikeda:134504,ikeda:054517}. 
Recent observations of magnetic order coexisting with 
the HFSC phase point towards a more complex situation and demands 
the reexamination and extension
of the basic picture~\cite{young2007,kenzelmann2008}. 
Neutron scattering measurements show that
the wave vector of AFM order is incommensurate,  
$\Q = \Qaf \pm \qiv$ with $\Qaf = (\pi,\pi,\pi)$ and
$\qiv \sim (0.12\pi,\pm 0.12\pi,0)$, and independent
of the orientation of inplane magnetic 
field.~\cite{kenzelmann2008,kenzelmann2010}
 The ordered Ce-moments $\vec{M}_{\rm AF}$ are oriented along the {\it c}-axis. 
 For the magnetic field along $[110]$, 
$\qiv$ is perpendicular to the applied magnetic field 
$\qiv \sim (0.12\pi,- 0.12\pi,0)$.
The data for the magnetic field along [100] direction
may be best interpreted as being due to two degenerate 
incommensurate AFM states with $\qiv \sim (0.12\pi,0.12\pi,0)$ 
and $\qiv \sim (0.12\pi,-0.12\pi,0)$,~\cite{kenzelmann2008,kenzelmann2010}  
as we will discuss below.

We may conclude that this magnetic order is stabilized by superconductivity 
because it is absent in the normal state. This strong cooperation between the magnetism 
and superconductivity contrasts the behavior of other heavy fermion 
superconductors where the magnetic order competes usually with 
superconductivity.~\cite{JPSJ_heavy}
It is worth noting that the HFSC phase is enlarged by 
pressure.~\cite{miclea2006} 
This feature is also different from the other magnetic phase diagram 
in Ce-based heavy fermion systems.~\cite{JPSJ_heavy} 

The observation of this unconventional magnetic order initiated 
a number of attempts 
for theoretical explanations \cite{yanase_LT,
yanase_JPSJ2009,miyake2008,aperis2008,aperis2010,agterberg2009,
ikedaAF,suzuki2010,vekhter2011}. 
Here we propose a theory based on the presence of an AFM quantum critical point 
near the superconducting phase of 
\Cof~\cite{paglione2003,bianchi2003,ronning2005,izawa2007,panarin2009}. 
As we have shown previously, AFM order appears within the inhomogeneous 
Larkin-Ovchinnikov state in the vicinity of 
the AFM quantum critical point.~\cite{yanase_LT,yanase_JPSJ2009} 
The mechanism coupling magnetism and FFLO superconductivity 
is consistent with the fact
that the magnetic order is restricted to the
superconducting phase in the $H$-$T$ phase 
diagram.~\cite{young2007,kenzelmann2008} 
Alternative proposals are based  on the 
emergence of a pair density wave state together with magnetic order 
in the HFSC phase.~\cite{aperis2008,aperis2010,agterberg2009}

 In order to identify the HFSC phase of \Co it is important to find
 unambiguous ways to compare the proposed phase with experimental 
 findings. Here we intend to examine the AFM-FFLO state proposed by us
 previously. First, we show that several phases can appear in the AFM-FFLO state 
when the magnetic field is applied along [100] or [010] direction. 
 Second, we clarify the magnetic structure of each phases and discuss
the consistency with the experimental results of \Cof.

\section{Ginzburg-Landau theory}

Our theoretical analysis is based on a phenomenological 
Ginzburg-Landau model formulated on the basis of the microscopic 
calculation in Refs. 29 and 42. 
and motivated by recent neutron scattering measurements 
of Kenzelmann \etal\cite{kenzelmann2008,kenzelmann2010} 
 We describe the AFM order in the inhomogeneous 
Larkin-Ovchinnikov state by means of the Ginzburg-Landau 
functional of the free energy,  
\begin{eqnarray}
\label{eq:GL}
  && \hspace*{-12mm}
\frac{F(\eta_1, \eta_2)}{F_0} = 
[(T/T_{\rm N}^{0} -1) + \xi_{\rm AF}^{2}(\qa - \qia)^{2}] \eta_1^2
\nonumber \\ && \hspace*{+5mm} 
+[(T/T_{\rm N}^{0} -1) + \xi_{\rm AF}^{2}(\qb - \qib)^{2}] \eta_2^2
\nonumber \\ && \hspace*{+5mm} 
+ \frac{1}{2}  (\eta_1^2 + \eta_2^2)^{2}
+ b \eta_1^2 \eta_2^2 \nonumber \\ && \hspace*{+5mm} 
+ c_1 H_{\rm x} H_{\rm y} (\eta_1^2 - \eta_2^2) 
\nonumber \\ && \hspace*{+5mm} 
- \frac{1}{2} \eta_1 \eta_2 \sum_{n} c_{2}(n) \delta(\qa - \qb, 2 n \qf) . 
\end{eqnarray}
We use an order parameter with two components, $\eta_1$ and $\eta_2$
corresponding to the two degenerate AFM states with 
$\Q \sim \Qaf \pm \qia$ and $\Q \sim \Qaf \pm \qib$.
The wave vectors of incommensurate AFM state in the system with 
full translational symmetry are given by
$\qia = (0.125\pi,0.125\pi,0)$ and 
$\qib = (-0.125\pi,0.125\pi,0)$. 
 In our study the incommensurate wave vectors $\qia$ and $\qib$ are not 
microscopically derived but assumed on the basis of the experimental 
results in Refs.~25 and 26. 
 We think that the $\qia$ and $\qib$ are pinned through nesting 
features in the band structures. 
Note that the order parameter $(\eta_1, \eta_2)$ is not a vector, 
but a director.

We describe the magnetic moment $M(\r) = e^{i \Qaf \cdot \r} M_{\rm AF}(\r)$
where $M_{\rm AF}(\r)$ is a slowly varying amplitude of the magnetic 
moment perpendicular to the applied magnetic field at $\r = (x,y,z)$.  
Then, the AFM staggered moment is given by
\begin{eqnarray}
 \label{eq:moment}
  && \hspace*{-5mm}
M_{\rm AF}(\r) = 
M_0 [\eta_1 \cos(\qa \cdot \r) + \eta_2 \cos(\qb \cdot \r)]. 
\end{eqnarray}
Both $M_0$ and $F_0$ are scaling factors to write
the Ginzburg-Landau free energy  in a dimensionless form 
(eq.(\ref{eq:GL})). 
Moreover, $T_{\rm N}^{0}$ is the ''bare'' \neel temperature 
(for $c_1 = c_{2}(n) = 0$, i.e. unrenormalized by
coupling to the magnetic field and to the FFLO-modulation 
of the superconducting phase), 
and $\xi_{\rm AF}$ is the correlation length of the AFM order. 
The inplane lattice constant is chosen as the unit length. 

Choosing the coupling constant $b$ positive we stabilize the 
''single-$q$'' magnetic structure $(\eta_1, \eta_2) \propto (1,0)$ or $(0,1)$, while 
a negative value for $b$ would favor the ''double-$q$'' magnetic structure 
$(\eta_1, \eta_2) \propto (1,1)$. From the microscopic point of view
$b > 0$ is more appropriate, since the single-$q$ magnetic 
phase gains more condensation energy than the double-$q$ 
phase in most cases. 
 We assume $b >0$ in this paper, while the cases of negative $b$ 
have been discussed in Ref.~43.

The constant $c_{1}$ denotes the coupling strength of 
the order parameter to the external magnetic 
field $\vH = (H_{\rm x},H_{\rm y},H_{\rm z})$.  
Comparing with  the neutron scattering data 
for $\vH \parallel [110]$,~\cite{kenzelmann2008} 
$c_1$ must be positive in \Cof, as to pick the incommensurate 
wave vector $\qib \perp [1\bar{1}0] $. 
This choice is consistent with our theoretical analysis 
for the AFM-FFLO state,~\cite{yanase_LT,yanase_JPSJ2009,yanase_JPCM} 
while it is incompatible with the theoretical scenario 
without taking the FFLO state into account.~\cite{suzuki2010,vekhter2011} 
When we consider the magnetic field along the [100] (or [010]) direction, 
this term is inactive and, therefore, the degeneracy of $\eta_1$ 
and $\eta_2$ remains. We focus on this situation in this paper.

The effects of the broken translational symmetry in the inhomogeneous 
Larkin-Ovchinnikov state are taken into account 
in the last term (commensurate term) of eq.(\ref{eq:GL}). 
 We define $\delta(\qa - \qb, 2 n \qf) = 1$ 
when the commensurate condition $\qa - \qb = 2 n \qf$ is satisfied 
for an integer $n$ and otherwise $\delta(\qa - \qb, 2 n \qf) = 0$. 
 The modulation vector of FFLO state is denoted as $\qf$, 
and then the order parameter of superconductivity is described as
$\Delta(\r) = \Delta_0 \sin(\qf \cdot \r)$. 
 The commensurate term describes the pinning effect of FFLO nodal planes 
for the AFM moment. 
 According to the microscopic calculation based on 
the Bogoliubov-de-Gennes (BdG) equation, 
the AFM moment is enhanced around the FFLO nodal planes where 
the superconducting order parameter 
vanishes.~\cite{yanase_LT,yanase_JPSJ2009,yanase_JPCM}  
 Then, we obtain the positive coupling constant $c_2(n) \geq 0$.

\section{Phase Diagram}

Let us now consider the FFLO state in a case with $2n\qf$ 
is close to $\qia - \qib$ for one integer $ n = N $, i.e. 
we assume  $\qia - \qib \sim 2 N \qf$. This has consequences 
for the structure of the magnetic order. 
When the wave vector $\qia - \qib$ is not in the vicinity of 
$2 n\qf$ for any integer $n$, the single-$q$ magnetic phase with 
$(\eta_1, \eta_2) \propto (1,0)$ or $(0,1)$ is realized.

 We determine the magnetic structure by minimizing the 
Ginzburg-Landau free energy (eq.(\ref{eq:GL})) with respect to 
the order parameters $\eta_1$ and $\eta_2$ and their momentum 
$\qa$ and $\qb$ for given temperature $T$ and FFLO wave vector $\qf$. 
We assume $\eta_1 \geq \eta_2 \geq 0$ without  loss of 
the generality. 
As for the direction of $\qf$, we assume 
$\qf = \qfs \hat{x}$ for $\vH \parallel [100]$ as 
in Refs.~29 and 42. 
 Then, the phase diagram is determined by 
the renormalized parameters $T/T_{\rm N}^{0}$ and $\xi_{\rm AF}q_0$,
where $q_0$ is defined by 
$\qia - \qib + 2 q_0 \hat{x} = 2 N \qf$. 
 The parameter $q_0$ describes the mismatch of the harmonic 
FFLO wave vector $2 N \qf$ and the incommensurability along 
the $\hat{x}$-axis $\qia -\qib$. 
 Note that $q_0 = 0$ when the condition 
$\qia - \qib = 2 N \qf$ is satisfied. 
 Since the FFLO wave number $\qfs$ grows with increasing 
magnetic field and/or decreasing temperature,~\cite{matsuda2007} 
$q_0$ increases with the magnetic field, and changes its sign at the 
line in the $H$-$T$ phase diagram.

 We find two possible phases. One is the single-$q$ phase 
where 
\begin{eqnarray}
 \label{eq:single-q}
  && \hspace*{-5mm}
\eta_1 > \eta_2,
\\ && \hspace*{-5mm}
 \label{eq:single-q1}
\qa = \qia + (1-x) q_0 \hat{x},
\\ && \hspace*{-5mm}
 \label{eq:single-q2}
\qb = \qib - (1+x) q_0 \hat{x}.
\end{eqnarray}
Note that $x$ lies between 0 and 1 and is determined by minimizing the free energy. 
The other is the double-$q$ phase where 
\begin{eqnarray}
 \label{eq:double-q}
  && \hspace*{-10mm}
\eta_1 = \eta_2,
\\ && \hspace*{-10mm}
 \label{eq:double-q1}
\qa = \qia + q_0 \hat{x},
\\ && \hspace*{-10mm}
 \label{eq:double-q2}
\qb = \qib - q_0 \hat{x}. 
\end{eqnarray}
 By analyzing the quadratic terms of eq.(\ref{eq:GL}), 
we find that the double-$q$ phase 
is stabilized immediately below the \neel temperature for 
$(\xi_{\rm AF}q_0)^{2} \leq c_2(N)/8$, while a 
single-$q$ phase 
is stabilized for $(\xi_{\rm AF}q_0)^{2} > c_2(N)/8$.

\begin{figure}[ht]
\begin{center}
\includegraphics[width=7cm]{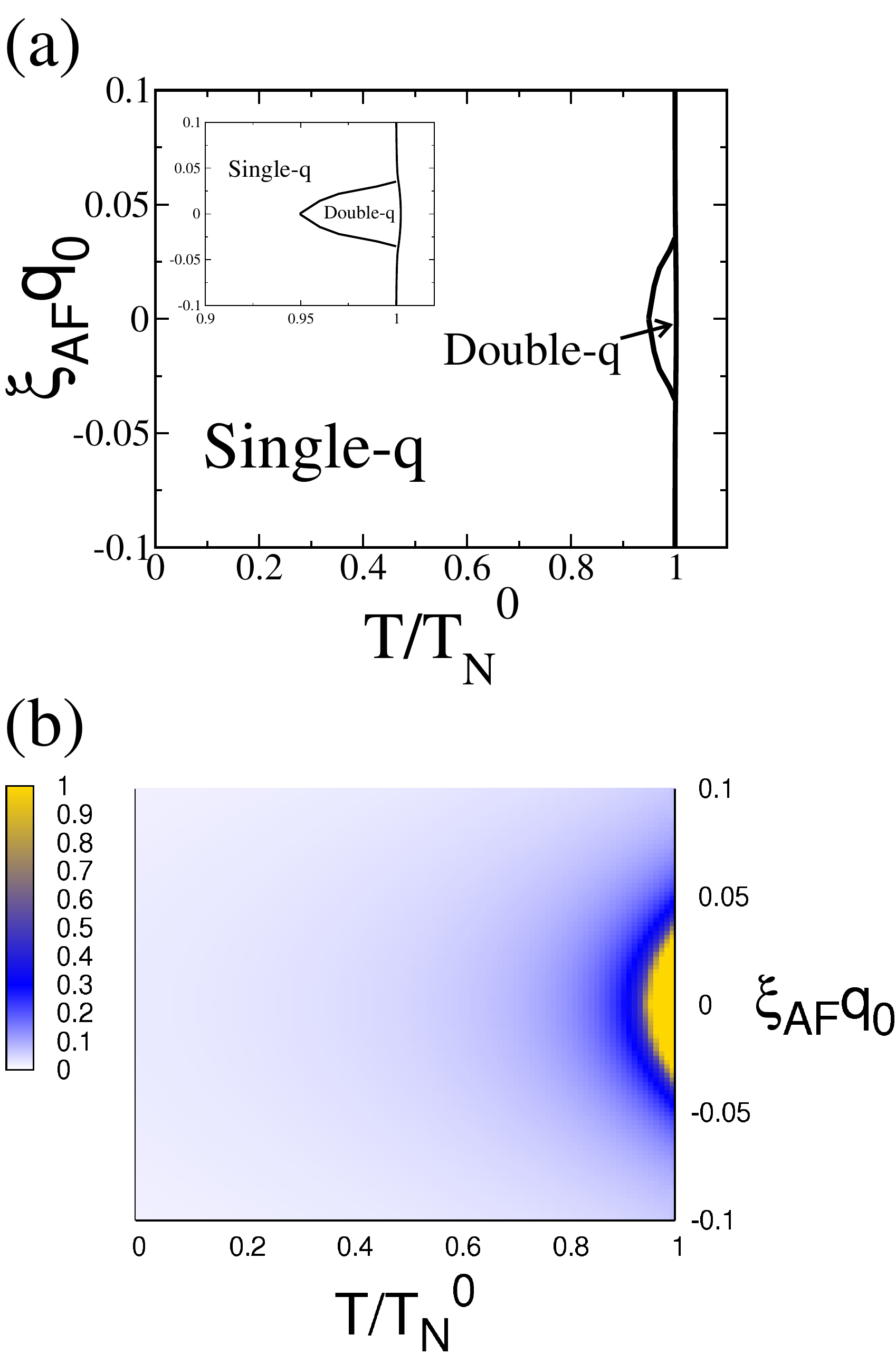}\hspace{1pc}%
\caption{(Color online)
(a) Phase diagram of Ginzburg-Landau model in eq.(\ref{eq:GL}) 
for $\xi_{\rm AF}q_0$ and the renormalized temperature 
$T/T_{\rm N}^{0}$. 
The definition of $q_0$ is given in the text. 
The $q_0$ increases with the magnetic field. 
The inset shows the same phase diagram scaled up around 
the \neel temperature. 
(b) The ratio of order parameters $\eta_2/\eta_1$. 
}
\end{center} 
\end{figure}

 Figure~1(a) shows the phase diagram obtained by the numerical 
minimization of the free energy in eq.(\ref{eq:GL}) 
using the parameters $b=0.1$ and $c_2(N) =0.01$. 
The single-$q$ phase is stable in most parameter regimes, 
because the quartic term $b \eta_1^2 \eta_2^2$ dominates. 
Only near the \neel temperature quadratic terms are leading and
can stabilize the double-$q$ phase for small 
$|\xi_{\rm AF}q_0| \leq \sqrt{c_2(N)/8}$. 
 As shown in the inset of Fig.~1(a), the \neel temperature 
increases slightly around the commensurate line $\xi_{\rm AF}q_0 =0$
due to the locking-in effect of the magnetic and the FFLO modulation. 
 However, the enhancement of the \neel temperature 
$T_{\rm N} - T_{\rm N}^{0} = \frac{c_2(N)}{4}  T_{\rm N}^{0}$ is rather small,
unless the coupling constant $c_2(N)$ is large. 
 Various phase diagrams for several parameter sets $(b,c(N))$ have been 
shown in Ref.~43. We here focus on the case of Fig.~1 and examine the 
consistency with experiments. 
 
 Figure~1(b) shows the ratio of order parameters $\eta_1$ and $\eta_2$. 
We find that $\eta_2/\eta_1 = 1$ in the double-$q$ phase while 
the ratio decreases in the single-$q$ phase 
with decreasing temperature $T/T_{N}^0$ 
and/or increasing the mismatch $|\xi_{\rm AF}q_0|$.

\begin{figure}[ht]
\begin{center}
\includegraphics[width=8cm]{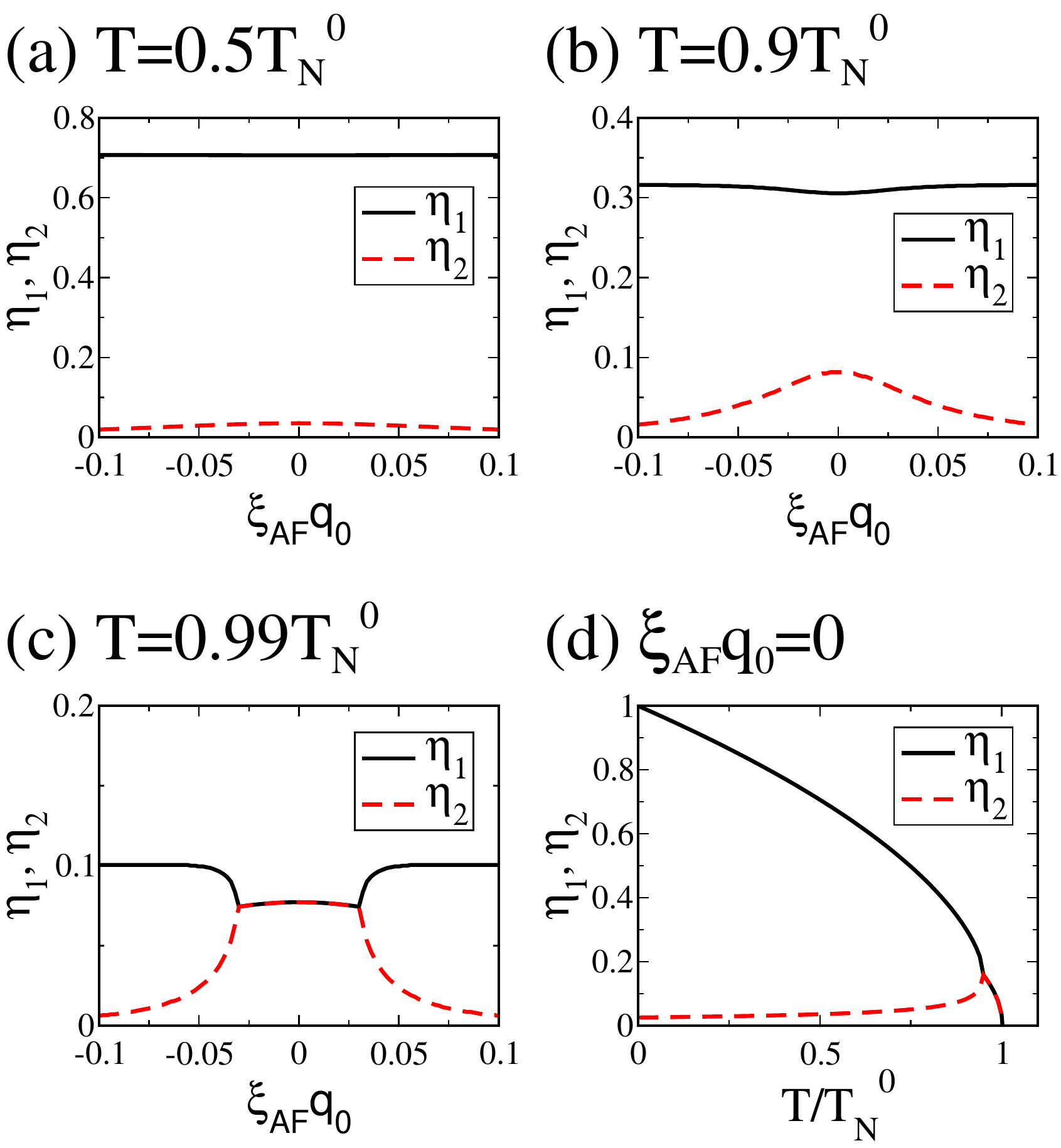}\hspace{1pc}%
\caption{(Color online)
The $\xi_{\rm AF}q_0$ dependences of order parameters 
$\eta_1$ and $\eta_2$ at (a) $T=0.5T_{\rm N}^{0}$, 
(b) $T=0.9T_{\rm N}^{0}$, and (c) $T=0.99T_{\rm N}^{0}$, 
respectively. 
(d) Temperature dependence of $\eta_1$ and $\eta_2$ at 
$\xi_{\rm AF}q_0 =0$. 
}
\end{center} 
\end{figure}

The behavior of the order parameter can be seen in the
plot of the $\xi_{\rm AF}q_0$ dependence of order parameters 
for $T = 0.5 T_{\rm N}^{0}$, $T = 0.9 T_{\rm N}^{0}$, and 
$T = 0.99 T_{\rm N}^{0}$ in Figs.~2(a)-2(c), respectively 
while the temperature dependence for $\xi_{\rm AF}q_0 =0$ is shown 
in Fig.~2(d). 
 The phase transition between the double-$q$ phase and single-$q$ phase 
is a continuous second-order transition. 
 This phase transition is characterized by the broken mirror symmetry 
with respect to the $x$- and $y$-axes in the single-$q$ phase.

\begin{figure}[ht]
\begin{center}
\includegraphics[width=8.5cm]{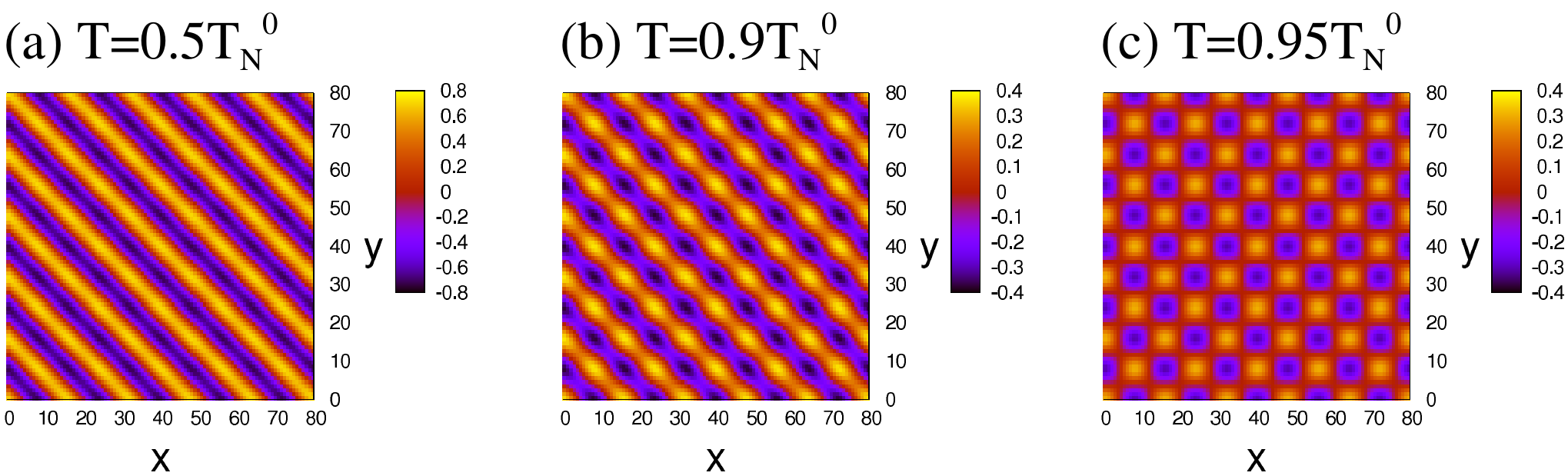}\hspace{1pc}%
\caption{(Color online)
The spatial dependences of AFM staggered moment 
$M_{\rm AF}(\r)$ at $\r = (x,y,0)$. 
(a) Single-$q$ phase at $T=0.5T_{\rm N}^{0}$, 
(b) Single-$q$ phase at $T=0.9T_{\rm N}^{0}$, 
and (c) Double-$q$ phase at $T=0.95T_{\rm N}^{0}$, 
respectively. 
We choose $\xi_{\rm AF}q_0 =0$. 
}
\end{center} 
\label{fig:stag-mom}
\end{figure}

 We show the spatial dependences of the AFM staggered moment 
$M_{\rm AF}(\r)$ for several parameters in Fig.~3. 
 Since the AFM moment is uniform along the $z$-direction, we 
show the $M_{\rm AF}(\r)$ in the $x$-$y$ plane. 
 The checkerboard magnetic structure is realized in the 
double-$q$ phase (Fig.~3(c)), while it changes to 
the stripe magnetic structure in the single-$q$ phase 
(Figs.~3(a) and 3(b)). 
 We investigate these phases in more details and 
discuss the experimental results of \Co in the next section.

\section{HFSC Phase in \Cof}

 In order to discuss the possible AFM-FFLO states in \Co on the basis of 
our Ginzburg-Landau analysis, we 
show the schematic phase diagram of \Co for the magnetic field 
and temperature in Fig.~4. 
 The amplitude of FFLO modulation vector $\qfs$ has
a maximal possible value of $\qfs \sim 1/\xi$ with $\xi$ being the 
coherence length of the superconducting state. 
 According to the experimental estimate of $\xi$, the minimum number 
of $N$, which satisfies the condition 
$\qia - \qib = 2 N \qf$ for commensurate wave vectors, 
is approximately $N \sim 4$. 
 This condition is satisfied in the FFLO state 
on a sequence of commensurate lines (dashed lines in Fig.~4) with 
$N = 4, 5, 6, 7 ....$, and the double-$q$ phase is stabilized 
around the AFM transition line (shaded area of Fig.~4). 
 Note that the double-$q$ phase does not appear 
close to the first-order normal-to-FFLO transition line 
at which the AFM moment as well as the superconducting order parameter 
appear discontinuously. 
 Since the coupling constant $c_2(N)$ decreases with increasing $N$, 
the double-$q$ phase is suppressed in the low magnetic field region.

\begin{figure}[ht]
\begin{center}
\includegraphics[width=5.5cm]{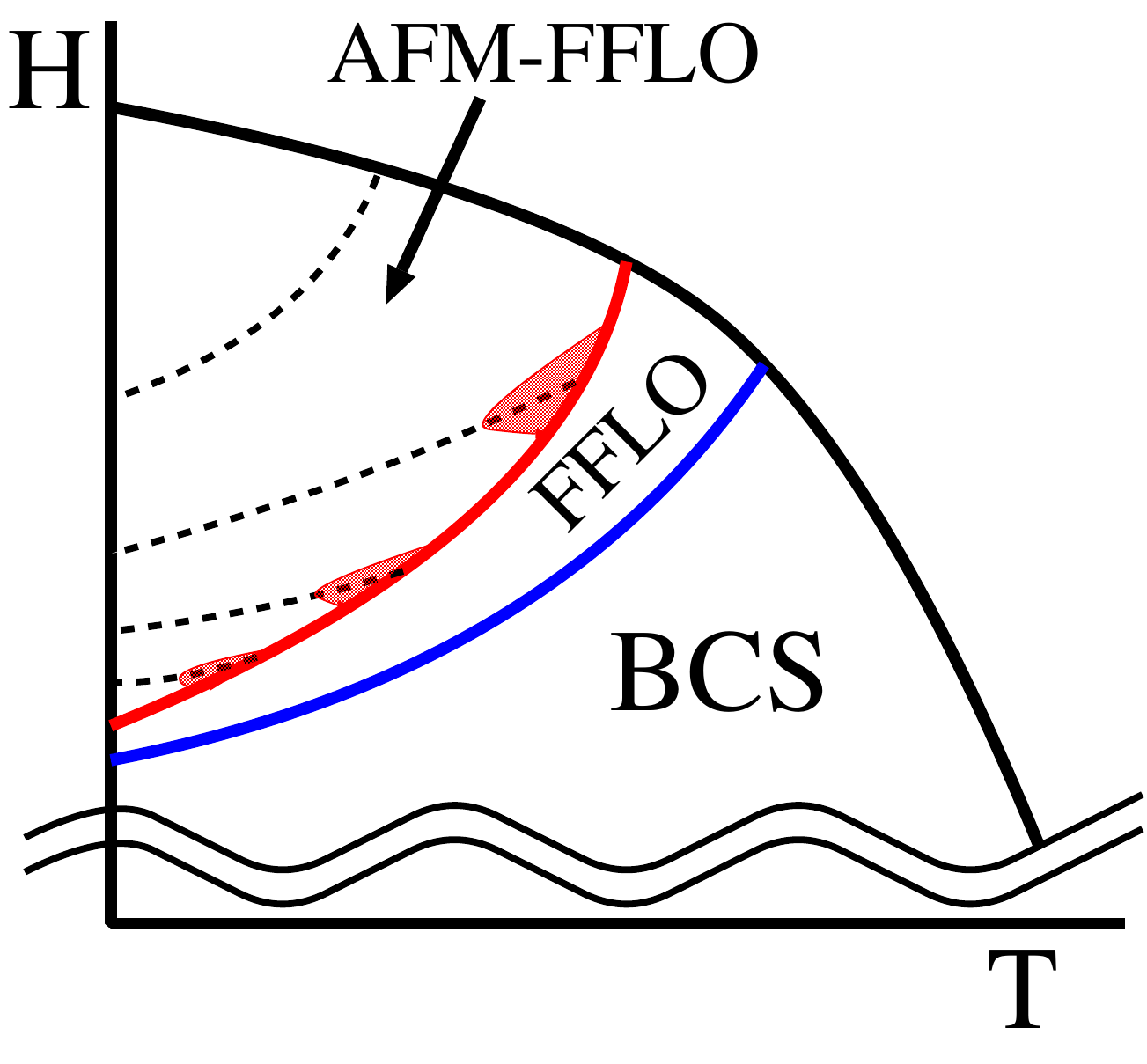}\hspace{1pc}%
\caption{(Color online)
Schematic phase diagram for the magnetic field along 
[100] direction. 
``BCS'', ``FFLO'', and ``AFM-FFLO'' states are shown in the figure. 
The dashed lines show the commensurate lines which are 
explained in the text. 
The shaded region shows the double-$q$ phase in the AFM-FFLO state. 
}
\end{center} 
\end{figure}

 It should be noticed that the AFM-FFLO phase is mostly covered by the 
single-$q$ phase. 
 Therefore, the experimental results of \Co should be compared with  
the properties of the single-$q$ phase. There are ways to distinguish
the single-$q$ phase experimentally 
from the double-$q$ phase. Indeed the experimental results 
are consistent with the single-$q$ phase, as we will discuss below.

 When we take into account the broken translational symmetry arising from 
the vortex lattice, other commensurate lines can appear in the 
phase diagram and stabilize the double-$q$ phase around $T=T_{\rm N}$. 
 However, it is expected that the effect of the vortex lattice 
on the magnetic order is smaller than that of the FFLO nodal planes 
when the Maki parameter is large enough to make the lattice spacing 
of vortices much larger than the coherence length.

\subsection{Neutron scattering}

The neutron scattering measurements have well
determined the structure of AFM order and its magnetic field 
dependence.~\cite{kenzelmann2008,kenzelmann2010}
 For the magnetic field along [100] direction 
the elastic Bragg peaks appear at both $\Q = \Qaf \pm \qia$ 
and the symmetry related $\Q = \Qaf \pm \qib$.~\cite{kenzelmann2008,kenzelmann2010} 
At first sight, this result seems to be incompatible with the single-$q$ phase 
in which, for instance, the Bragg peaks at $\Q = \Qaf \pm \qib$ should be much weaker than
those at $\Q = \Qaf \pm \qia$ (see Fig.~5(a)). However, taking into account that the
sets of wave vectors correspond to degenerate single-$q$ phases in a field along [100], it is rather likely
that the domains have formed of the two states  $ |\eta_1| > | \eta_2| $ and $ |\eta_1| < | \eta_2| $.
Because the domain structure is history-dependent, it can be identified in future experiments, for example,
by changing the field orientation slightly away from [100] direction in order to lift the degeneracy.

\begin{figure}[ht]
\begin{center}
\includegraphics[width=8cm]{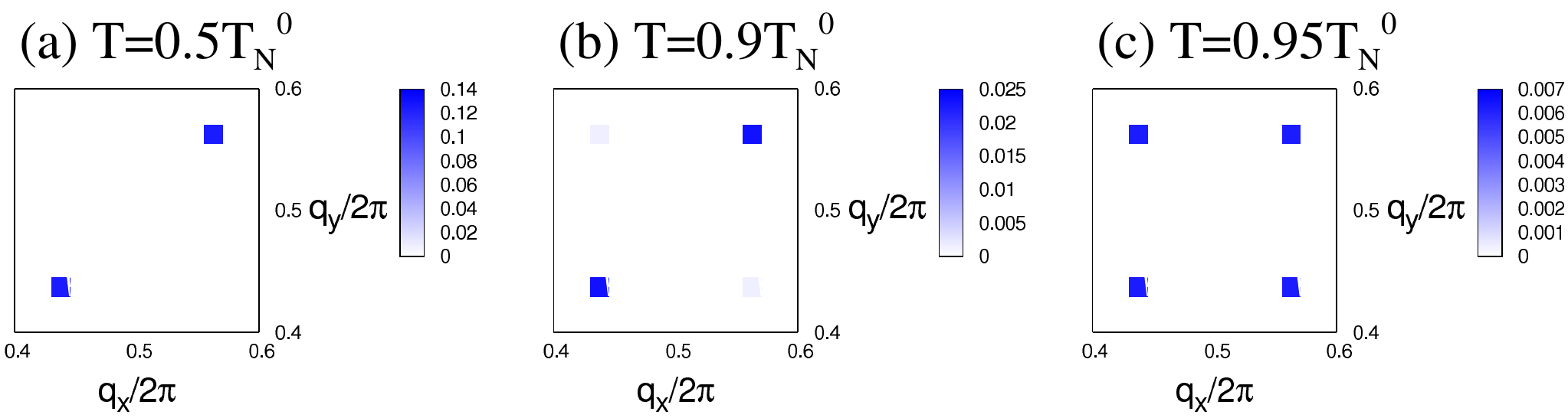}\hspace{1pc}%
\caption{(Color online)
The weight of Bragg peaks defined by $|M(\q\,)|^2$ with 
$M(\q\,) = \frac{1}{N_0} \sum_{\r} M(\r) e^{{\rm i} \q \cdot \r}$ 
and $\q = (q_{\rm x}, q_{\rm y},\pi)$. 
The summation $\sum_{\r}$ is taken over the $N_0$ lattice sites. 
(a) Single-$q$ phase at $T=0.5T_{\rm N}^{0}$, 
(b) Single-$q$ phase at $T=0.9T_{\rm N}^{0}$, 
and (c) Double-$q$ phase at $T=0.95T_{\rm N}^{0}$, 
respectively. 
We choose $\xi_{\rm AF}q_0 =0$. 
}
\end{center} 
\end{figure}

As discussed above and stated in eqs.(\ref{eq:single-q1}), (\ref{eq:single-q2}), 
(\ref{eq:double-q1}) and (\ref{eq:double-q2}), 
the incommensurate wave vectors $\qa$ and $\qb$ depend on the magnetic 
field. However, the deviation from $\qia$ is rather small for 
the main Bragg peak, with
$|\qa - \qia|  \leq \xi_{\rm AF}^{-1} \sqrt{c_2(N)/8}$. 
For the parameters $c_2(N) = 0.01$ and $\xi_{\rm AF} =3$, 
we obtain $|\qa - \qia|  < 0.004 \pi$.
 Moreover, the shift of $\qa$ rapidly decreases upon lowering the 
temperature below $T_{\rm N}$ (see Fig.~6). 
Therefore, additionally challenged by the tininess of the magnetic moment near the \neel temperature, 
most likely the shift of Bragg peaks is experimentally unobservable. 
Indeed, in the experiments no magnetic field dependence of 
$\qa$ has been observed so far.~\cite{kenzelmann2008,kenzelmann2010}. 
Nevertheless, 
it would be interesting to search for a possible shift of $\qa$ in future experiments.

\begin{figure}[ht]
\begin{center}
\includegraphics[width=6.5cm]{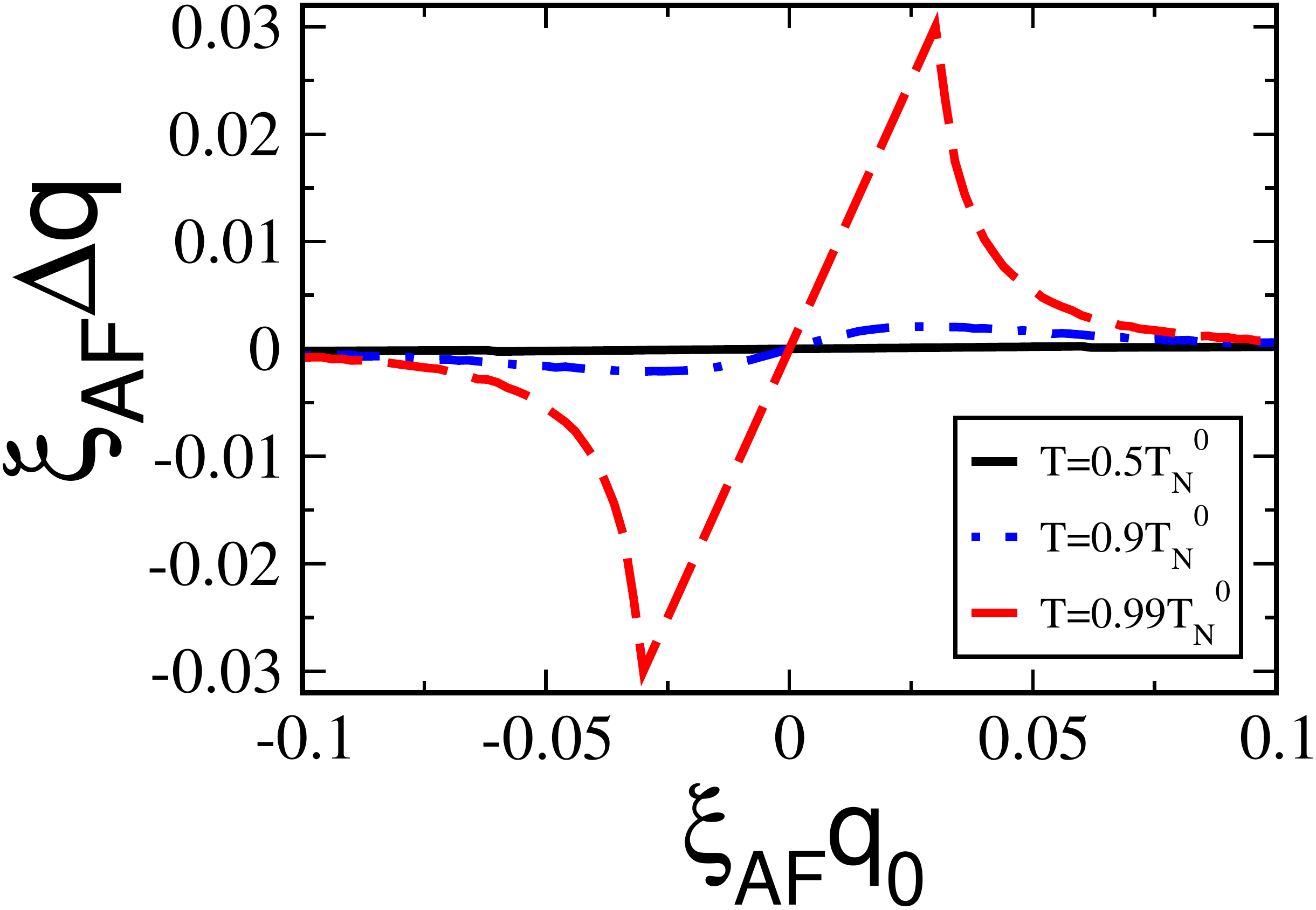}\hspace{1pc}%
\caption{(Color online)
The shift of the position of main Bragg peak from $\qia$. 
We plot $\xi_{\rm AF} \Delta q$ for $T=0.5T_{\rm N}^{0}$, 
$T=0.9T_{\rm N}^{0}$, and $T=0.99T_{\rm N}^{0}$, 
where $\Delta q$ is defined by $\Delta q \hat{x} = \qa - \qia$. 
}
\end{center} 
\end{figure}

Finally, we would like to state that also the case of $b <0$ stabilizing 
the double-$q$ phase shows four Bragg peaks consistent with the neutron 
scattering data for $\vH \parallel [100]$ without having to assume
domain formation.  
However, this case yields a sizable shift of $\qa$ at low temperature, 
and therefore seems to be incompatible with the neutron scattering 
data.~\cite{yanase_ICHE} 
Moreover, as we will elucidate now this scenario is incompatible 
with the NMR measurements.

\subsection{NMR}

 The first evidence for AFM order in the HFSC phase of 
\Co has been obtained by the NMR measurement.~\cite{young2007}
The analysis of the NMR spectrum pointed towards
incommensurate AFM order. This interpretation 
has subsequently been confirmed by the neutron scattering 
measurements in a decisive way.~\cite{kenzelmann2008,kenzelmann2010}
Recent analysis showed that the direction of AFM moment 
along $c$-axis is also consistent with the NMR data by assuming 
dipolar hyperfine 
coupling.~\cite{curro2009,mitrovic2008,koutroulakis2010,kumagaiprivate} 
 
Applying the magnetic field parallel to the $x$-axis there are
three distinct In-sites for NMR, the inplane In(1)-site and the out-of-plane
In(2a)- and In(2b)-site. The latter two lie on $yz$- and $zx$-planes, respectively, 
relative to the Ce-layers. Looking at the dipolar fields due to the
magnetic moments on the Ce-sites only the In(2b)-site develops
a field component parallel to the $x$-axis which contributes to the Knight
shift. Restricting to the two nearest Ce-ions of a In(2b)-site 
at $ \vec{r} $ we obtain for the dipolar field, 
\begin{equation}
H_x (\vec{r}) = A [ M(\vec{r}+\vec{b}_+) - M(\vec{r}+\vec{b}_-)], 
\end{equation}
with $A$ a constant and $ \vec{b}_{\pm} = (\pm 1/2,0,c') $ with 
$ c' $ the distance of In(2b) site from the Ce-plane. 
Note that there are In(2b)-sites above and below the Ce-plane. 
Using eq.(\ref{eq:moment}) we derive, 
\begin{eqnarray}
&& \hspace*{-15mm} H_x (\vec{r}) = 
\nonumber \\ && \hspace*{-15mm}
A' M_0 \cos (\vec{Q}_0 \cdot \vec{r}) [ \eta_1 \cos(\qa \cdot \vec{r}) + \eta_2 \cos(\qb \cdot \vec{r}) ],
\end{eqnarray}
which eventually leads to the field distribution $ P(h) $ observable in the Knight shift,
\begin{equation}
P(h) = \frac{1}{N_0} \sum_{\vec{r}} \delta(h - H_x(\vec{r}) ) \; .
\label{eq:internal_field}
\end{equation}
The sum runs over $ N_0 $ sites of square lattice 
representing the In(2b)-sites. 

\begin{figure}[ht]
\begin{center}
\includegraphics[width=8cm]{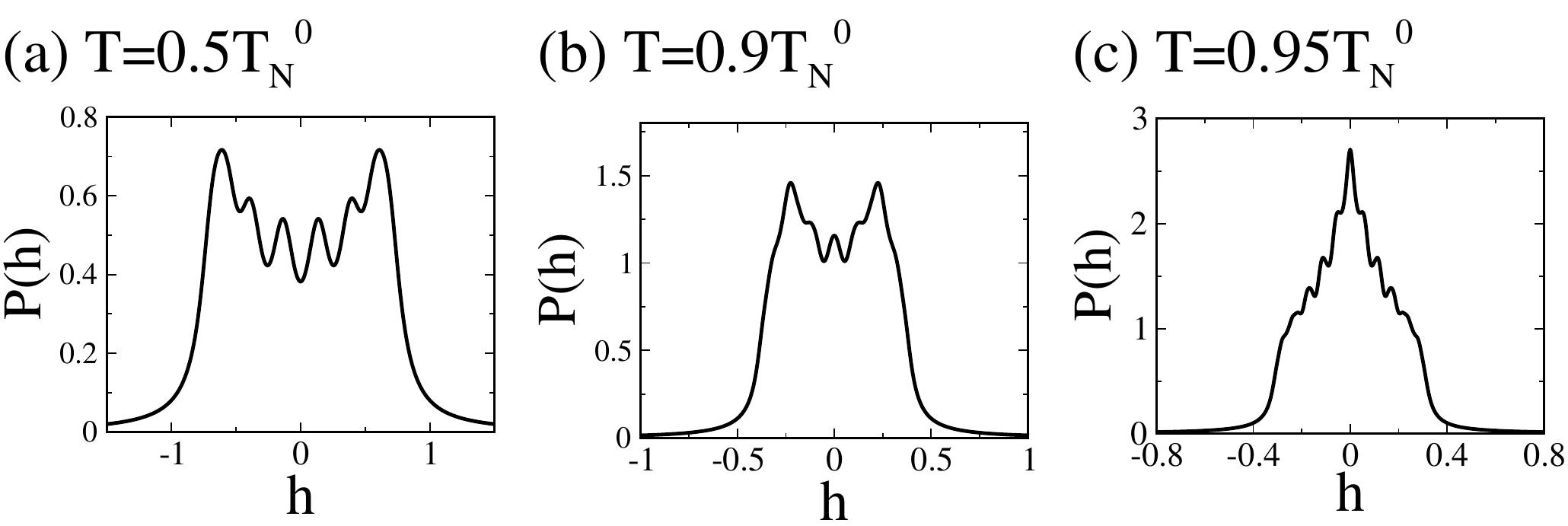}\hspace{1pc}%
\caption{
The distribution function of the internal field $P(h)$ 
at the In(2b) sites given by eq.(\ref{eq:internal_field}). 
The parameters are the same as in Figs.~\ref{fig:stag-mom}(a), 3(b) and 3(c), 
respectively. We choose $AM_{0} = \frac{1}{2}$. 
}
\label{fig:ph}
\end{center} 
\end{figure}

It is obvious that $ P(h) $ shows a broad distributions with two peaks at the upper and lower edge of the distribution, if $ (\eta_1,\eta_2) \propto (1,0) $ or $ (0,1) $ (single-$q$ phase as in Fig.~\ref{fig:ph}(a)). A single peak in the center of the distribution ($ h=0 $) is found for $ |\eta_1| = | \eta_2| $ (double-$q$ phase as in Fig.~\ref{fig:ph}(c)). In Fig.~\ref{fig:ph} we show the evolution of the distribution function $ P(h) $ for the same parameters as 
used for Figs.~\ref{fig:stag-mom}(a-c). 

The experimentally observed Knight shift distribution has the shape of a broad double peak structure. This is consistent with the realization of a single-$q$ phase and clearly does not fit to the double-$q$ phase. Thus, the phase diagram with the AFM-FFLO state, which is almost entirely covered by the single-$q$ phase, is consistent with the actual NMR measurements. 

 We now turn to the relationship between the analysis given in this paper 
and our previous microscopic study of the spatial structure of AFM-FFLO state. 
We have shown that three basic forms 
of the AFM-FFLO state are possible. \cite{yanase_JPCM}
We distinguish (1) the {\it 'extended'} case for which the magnetic moments are slightly
suppressed around the nodal planes of FFLO-superconducting order parameter, 
(2) the {\it 'weakly localized'} case in which the magnetic moments are slightly enhanced
around the nodal planes, and (3) the {\it 'strongly localized'} case with a 
magnetic order parameter very localized around the nodal planes. 
Our phenomenological theory in this paper is justified in the cases 
(1) and (2), while it is invalid for the case (3) because 
the incommensurate wave vector $\qiv$ would not be parallel 
to $[110]$ and $[1\bar{1}0]$ for the 
magnetic field along [100] axis. 
According to our analysis, 
only the case (2) is consistent with the 
neutron scattering measurements in which the position of magnetic Bragg 
peaks is independent of the magnetic field and $\qiv$ is perpendicular 
to the field ($\qiv \perp \vec{H}$) for $\vec{H} \parallel [110]$ or 
$\vec{H} \parallel [1\bar{1}0]$.~\cite{kenzelmann2008,kenzelmann2010} 

In the case (2), a weak enhancement of magnetic moment around the 
FFLO nodal planes, which is not taken into account in this paper, 
affects the NMR spectrum. This results in tails at the edges of 
the distribution function $P(h)$. This additional 
contribution to $P(h)$ does, however, not alter the qualitative feature of
the double peak structure anticipated for the single-$q$ phase.

 We here discuss the NMR measurements at In(1)- and In(2a)-sites. 
Since the AFM moment does not (or only weakly) yield internal field at  
these sites, we may avoid the effect of AFM order and elucidate 
the magnetic properties of FFLO state without AFM order. 
Recent NMR measurement at In(1)- and In(2a)-sites actually reports an 
evidence for the FFLO superconducting state in \Cof .~\cite{kumagaiprivate}

\subsection{Ultrasonic measurement}

As we have shown in Fig.~4, the region of the AFM-FFLO state of \Co is dominantly
the single-$q$ phase. 
 Because the order of single-$q$ phase has $ Z_2 $ character, a 
 probe sensitive to the crystal symmetry could give further helpful evidence for this phase.
Ultrasound measurements provide such an experimental tool, as we discuss
here. 

AFM order and crystal lattice deformations are coupled to each other through
the band structure and spin-orbit coupling. Avoiding microscopic details, 
we may describe this interplay
by means of a general Ginzburg-Landau free energy, 
\begin{eqnarray}
\label{eq:ex-GL}
&& \hspace*{-12mm}
F' =  F(\eta_1, \eta_2) + F_{\small \mbox{el-af}} + F_{\rm el}, 
\\ && \hspace*{-12mm} 
\label{eq:el-af}
F_{\small \mbox{el-af}} = 
\{\gamma_1 (\ep_{\rm xx} + \ep_{\rm yy}) + \gamma_2 \ep_{\rm zz}\} 
(\eta_1^2 + \eta_2^2) 
\nonumber \\ && \hspace*{3mm} 
+ \gamma_3 \ep_{\rm xy} (\eta_1^2 - \eta_2^2),  
\\ && \hspace*{-12mm} 
F_{\rm el} = \frac{1}{2} [
C_{11} (\ep_{\rm xx}^2 
+  \ep_{\rm yy}^2)
+ C_{33} \ep_{\rm zz}^2
\nonumber \\ && \hspace*{0mm} 
+ 2 C_{12} \ep_{\rm xx} \ep_{\rm yy}
+ 2 C_{13} ( \ep_{\rm xx} + \ep_{\rm yy}) \ep_{\rm zz}
\nonumber \\ && \hspace*{0mm} 
+ 4 C_{44} (\ep_{\rm yz}^2
+ \ep_{\rm xz}^2)
+ 4 C_{66} \ep_{\rm xy}^2 
], 
\end{eqnarray}
where the coefficients $C_{ij}$ represent elastic constants of a tetragonal crystal lattice, 
and $\ep_{ij} = \frac{1}{2} (\partial u_i/\partial x_j + \partial
u_j/\partial x_i)$ defines the strain tensor with $\vec{u}$ being the local 
displacement vector. $ \gamma_i $ are coupling constants between
strain and magnetic order parameter. 

We examine now the coupling to ultrasound modes, longitudinal (L) and transversal (T1,T2)
with inplane and {\it c}-axis propagation directions. 
For the propagation along [100] the following strain
fields are involved,
\begin{equation}
\mbox{L: } \epsilon_{xx}, \quad \mbox{T1: }  \epsilon_{xy}, \quad 
\mbox{T2: } \epsilon_{xz},
\end{equation}
while for [110],
\begin{equation}
\mbox{L: } \epsilon_{xx}=\epsilon_{yy}=\epsilon_{xy},  \mbox{  T1: }  \epsilon_{xx} = - \epsilon_{yy}, 
\mbox{  T2: } \epsilon_{xz}=\epsilon_{yz},
\end{equation} 
and for [001],
\begin{equation}
\mbox{L: } \epsilon_{zz}, \mbox{ T: } \epsilon_{xz}, \epsilon_{yz}.
\end{equation}
Note that for T1 (T2)  the transverse polarization lies in the basal plane (along the $z$-axis) and
for T it has two independent polarization directions in the basal plane. 

The sound velocity corresponding to these modes can be renormalized at the phase transition
to the AFM-FFLO phase. This corresponds to the renormalization of the elastic constants, given
by
\begin{equation}
\tilde{C}_{\alpha \beta} = \frac{d^2 F'}{d \epsilon_{\alpha} d \epsilon_{\beta}} = C_{\alpha \beta} + \sum_{n=1,2}
\frac{\partial^2 F'}{\partial \epsilon_{\alpha} \partial \eta_n}
\frac{\partial \eta_n}{\partial \epsilon_{\beta} } ,
\end{equation} 
at equilibrium, i.e. 
\begin{equation}
\frac{\partial F'}{\partial \eta_n} = \frac{\partial F'}{\partial \epsilon_{\alpha}} = 0 \; .
\end{equation}
We find coupling to the ultrasound mode, if the corresponding terms,
\begin{equation}
\left. \frac{\partial^2 F_{\small \mbox{el-af}}}{\partial \epsilon_{\alpha} \partial
 \eta_n} \right|_{\epsilon_{\alpha}=0; \eta_n = \eta_{n0}}, 
\end{equation}
are finite. We give now a list of couplings for the two phases, single-$q$ [$ (\eta_1,\eta_2) \propto (1,0) $] and double-$q$ [$ (\eta_1,\eta_2) \propto (1,1) $]. 

\begin{table}
\caption{Coupling of ultrasound modes to single- and double-$q$ phases: ''yes'' or ''no'' denote which elastic constants are
renormalized or unrenormalized.}
\label{table1}
\begin{center}
\begin{tabular}{lccc}
\hline
mode & single-$q$ & double-$q$ & $C_{ij} $ \\
\hline
{\bf [100]} & & & \\
L & yes & yes & $C_{11}$ \\
T1 & yes & yes &$ C_{66} $\\
T2 & no & no & $C_{44}$ \\
\hline
{\bf [110]} & & & \\
L & yes & yes & $C_{11} + C_{12} + 2 C_{66}$ \\
T1 & no & no & $C_{11} - C_{12}$ \\
T2 & no & no & $C_{44}$ \\
\hline
{\bf [001]} & & & \\
L & yes & yes & $C_{33} $\\
T & no & no & $C_{44}$ \\
\hline
\end{tabular}
\end{center}
\end{table}

\begin{table}
\caption{Anomaly of ultrasonic sound mode at the magnetic transition. 
The sound velocity (and elastic constant) shows a ``kink'', ``jump'', 
or ``divergence''. 
In case of the double transition, magnetic transition to the double-$q$ 
state occurs at $T=T_{\rm N}$ 
and that to the single-$q$ state occurs at $T=T_{\rm N2}$. 
In case of the single transition, the transition to 
the single-$q$ state occurs at $T=T_{\rm N}$. 
}
\label{table2}
\begin{center}
\begin{tabular}{lccc}
\hline
 & \multicolumn{2}{c}{double transition} & single transition \\
\cline{2-4}
mode  & $T=T_{\rm N}$ & $T=T_{\rm N2}$ & $T=T_{\rm N}$  \\
\hline
T1 for [100] & kink & divergence & jump \\
\hline
L for all & jump & jump & jump \\
\hline
\end{tabular}
\end{center}
\end{table}

Table \ref{table1} shows that both single- and double-$q$ phases couple 
in the same pattern to the different modes. 
Thus there are no selection rules distinguishing the two phases. 
Important is, however, to notice that both phases couple not only 
to longitudinal modes, but also 
to a transverse (T1) mode propagating in the [100] direction. 
This is a signature of the multi-component (magnetic) order parameter, 
since a single component magnetic order, such as the commensurate 
AFM order with $\Q = \Qaf$, does not couple to any transverse mode. 
 The anomaly in the T1 mode arises from the last term of 
eq.(\ref{eq:el-af}) which is described as 
$\gamma_3 \ep_{\rm xy} Q_{\rm xy}$ with use of the quadrupole order 
parameter $Q_{\rm xy} = \eta_1^2 - \eta_2^2$. 
 Thus, the internal degree of freedom having a quadrupole symmetry 
manifests itself in the transverse sound mode, as in the non-magnetic 
quadrupole order appearing in many heavy fermion systems.~\cite{quadrupole}

 We find characteristic behaviors of this sound mode and summarize in
Table \ref{table2}. 
 For $|\xi_{\rm AF}q_0| \leq \sqrt{c_2(N)/8}$ the double magnetic
transitions occur at $T=T_{\rm N}$ and $T=T_{\rm N2}$ 
($T_{\rm N2} < T_{\rm N}$). 
 Then, the T1 sound mode propagating along [100] direction shows a kink 
at $T=T_{\rm N}$ and diverges at $T=T_{\rm N2}$. 
 On the other hand, when the single magnetic transition to the single-q
phase occurs for $|\xi_{\rm AF}q_0| \geq \sqrt{c_2(N)/8}$, 
this sound mode shows a jump at $T=T_{\rm N}$. Thus, these two cases 
can be distinguished by the ultrasonic measurement for the T1 mode 
along [100] direction. 
All longitudinal modes show a jump at the magnetic phase transition.

 The ultrasonic measurement by Watanabe \etal actually found the 
anomaly in the T1 mode along [100] direction, but it has been 
attributed to the modulation of superconducting order parameter 
in the FFLO state.~\cite{watanabe2004} 
More detailed ultrasonic measurement with focus on the AFM order 
is highly desired to identify the HFSC phase of \Cof. 

It is obvious from eq.(\ref{eq:el-af}) that a uniaxial stress 
along $[110]$ or $[1\bar{1}0]$ direction ($ \epsilon_{xy} \neq 0 $) 
would split the degeneracy between $ \eta_1 $ and $ \eta_2 $ such that a
single domain phase could be obtained for the single-$q$ phase.

\section{Summary and Discussion}

In this study we investigated the coexistence of the incommensurate 
AFM order with the FFLO superconducting state to interpret the properties 
of the HFSC phase of \Cof. 
We assume that two incommensurate AFM modulations with $\qiv \parallel [110]$ and
$\qiv \parallel [1\bar{1}0]$ are degenerate for the
magnetic field along $[100]$ direction. Taking the coupling of the magnetic phase and 
the FFLO modulation of the superconducting phase into account we show that there
could be multiple AFM phases in  the $H$-$T$ phase diagram.
The so-called ''single-$q$'' phase corresponding to one of the two incommensurate
wave vectors $ \qiv $ is dominant in the phase diagram. However, the ''double-$q$'' phase 
superposing both wave vectors can be induced under certain conditions, if there is 
commensuration condition between FFLO and AFM modulations. This phase 
is unstable against the second order transition towards a single-$q$ state as our
Ginzburg-Landau model demonstrates. 

We have shown that the single-$q$ phase is consistent with the 
experimental results of the NMR and neutron scattering measurements 
by taking into account the domain structure of two degenerate 
single-$q$ phases. 
A future experiment by the ultrasonic measurement has been discussed. 
  
 Finally, we discuss the field orientation dependence of the AFM order 
in \Cof. 
 Since the staggered magnetic moment lies along the $c$-axis 
owing to the spin-orbit coupling, the AFM order should be suppressed 
by the magnetic field along $[001]$ direction. 
 Furthermore, a small $H_{\rm c2}$ along $c$-axis indicates a weak 
paramagnetic effect which is unfavorable for the magnetic 
order.~\cite{ikedaAF,suzuki2010} 
 Thus, it is expected that the AFM order does not occur for this field 
direction even when the FFLO superconducting state is stabilized. 
 Indeed, recent experiments have shown that the magnetic order is 
suppressed for $\vec{H} \parallel [001]$.~\cite{blackburn2010,paulsen2011}

\section*{Acknowledgements}

 The authors are grateful to D. Aoki, D. F. Agterberg, J. P. Brison, 
N. J. Curro, J. Flouquet, S. Gerber, R. Ikeda, M. Kenzelmann, 
G. Knebel, K. Kumagai, K. Machida, Y. Matsuda, V. F. Mitrovi\'c, 
K. Mitsumoto, and H. Tsunetsugu 
for fruitful discussions. 
 This work was supported by 
a Grant-in-Aid for Scientific Research on Innovative Areas
``Heavy Electrons'' (No. 21102506) from MEXT, Japan. 
 It was also supported by a Grant-in-Aid for 
Young Scientists (B) (No. 20740187) from JSPS. 
 Numerical computation in this work was carried out 
at the Yukawa Institute Computer Facility. 
YY is grateful for the hospitality of the Pauli Center of ETH Zurich. 
This work was also supported by the Swiss Nationalfonds 
and the NCCR MaNEP.

\bibliographystyle{jpsj}
\bibliography{FFLO_GL}

\end{document}